\documentclass[aps,preprint,prd,showpacs,nofootinbib]{revtex4}

\usepackage{latexsym}
\usepackage{amsmath}
\usepackage{graphicx}
\usepackage{subfigure}
\usepackage{dcolumn}
\usepackage{bm}
\usepackage{amssymb}
\usepackage{latexsym}
\usepackage{ulem}
\usepackage{color}
\usepackage[colorlinks,linkcolor=magenta,anchorcolor=cyan,citecolor=blue]{hyperref}

\def\be{\begin{equation}}
\def\ee{\end{equation}}
\def\ba{\begin{eqnarray}}
\def\ea{\end{eqnarray}}

\def\lf{\left}
\def\rt{\right}

\begin{document}

\title{ Phase-transition sound of inflation at gravitational waves detectors}

\author{Yu-Tong Wang$^{1}$\footnote{wangyutong@ucas.ac.cn}}
\author{Yong Cai$^{1,2}$\footnote{caiyong13@mails.ucas.ac.cn}}
\author{Yun-Song Piao$^{1,3}$\footnote{yspiao@ucas.ac.cn}}

\affiliation{$^1$ School of Physics, University of Chinese Academy of
Sciences, Beijing 100049, China}

\affiliation{$^2$ Department of Physics and Astronomy, University
of Pennsylvania, Philadelphia, Pennsylvania 19104, USA }

\affiliation{$^3$ Institute of Theoretical Physics, Chinese
Academy of Sciences, P.O. Box 2735, Beijing 100190, China}

\begin{abstract}

It is well-known that the first-order phase transition (PT) will
yield a stochastic gravitational waves (GWs) background with a
peculiar spectrum. However, we show that when such a PT happened
during the primordial inflation, the GWs spectrum brought by the
PT will be reddened, which thus records the unique voiceprint of
inflation. We assess the abilities of the GW detectors to
detect the corresponding signal.

\end{abstract}

\maketitle


\section{Introduction}

Inflation is the current paradigm of early universe
\cite{Guth:1980zm}\cite{Linde:1981mu}\cite{Albrecht:1982wi}\cite{Starobinsky:1980te}.
It predicts nearly scale-invariant scalar perturbation, which is
consistent with the cosmic microwave background (CMB)
observations, as well as the gravitational waves (GWs).

The primordial GWs background spans a broad frequency-band,
$10^{-18}-10^{10}$ Hz,
e.g., \cite{Lasky:2015lej}\cite{Wang:2016tbj}. It is usually thought
that the discovery of primordial GWs will solidify our confidence
that inflation has ever occurred. The primordial GWs at ultra-low
frequency $10^{-18}-10^{-16}$ Hz may induce the B-mode
polarization in the CMB
\cite{Kamionkowski:1996ks}\cite{Kamionkowski:1996zd}.
Recently, the combination of Planck, BICEP and Keck Array's 95 GHz
data has put the constraint on the tensor-to-scalar ratio to
$r_{0.05}<0.07$ (95\% C.L.) \cite{Array:2015xqh}. However, if
$r\ll 0.001$, detecting the primordial GWs and verifying the
inflation with CMB will be extremely difficult.

Recently, the LIGO/VIRGO Scientific Collaboration, using the laser
interferometer, has observed the GWs signals consistent with the
binary black holes coalescence
\cite{Abbott:2016blz}\cite{TheLIGOScientific:2016wyq}. This opens
a new window for exploring the inflation. It is well-known that the
stochastic GWs background predicted by the slow-roll inflation is
too small to be detected by the GWs detectors, such as Advanced
LIGO and LISA, see \cite{Guzzetti:2016mkm}\cite{Bartolo:2016ami}
for reviews on other possible sources. How can we verify the
inflation with the GWs detectors?

The initial state of sub-horizon GWs mode $\gamma_k$ in de Sitter space
is $\gamma_k\sim {1\over a\sqrt{2k}}e^{-ik\tau}$, i.e., the
Bunch-Davis (BD) vacuum, where the conformal time $\tau=\int
dt/a$. Its power spectrum is $P_T({k\gg aH})\sim k^2$. However,
the inflation will stretch the sub-horizon GWs mode outside the
horizon. When $\gamma_k$ is leaving the horizon ($k=aH$),
$\gamma_k\sim {H\over \sqrt{2k^3}}e^{-ik\tau}$, hereafter
$\gamma_k$ becomes superhorizon and we have $P_T\sim k^0$. Thus one
significant effect of inflation (nearly exponential expansion) is
that it reddens the sub-horizon spectrum $P_T({k\gg aH})$. Could
we find this reddening effect of inflation with the GWs detectors? It is possible, if a \textit{known} sub-horizon GWs background,
stronger than the BD state, was produced at some time
(corresponding to the wavebands of GWs detectors) during
inflation.

Phase transition (PT) is ubiquitous in the early universe. In this
Letter, we find that when the first-order PT happened during
inflation, the GWs spectrum brought by the PT will be reddened,
which thus records the unique voiceprint of inflation.
Especially, we show that Advanced LIGO and LISA have abilities to
detect the corresponding signals. It should be mentioned
that Ref.\cite{Jiang:2015qor} discussed the detectability of such
a signal in light of CMB experiments, see also
\cite{Jinno:2011sw}. Our result suggest a novel possibility of
witnessing inflation, which has not been noticed before.

\section{Inflation and primordial PT GWs}

Historically, it is the first-order PT in particle physics that
has motivated the idea of (old) inflation \cite{Guth:1980zm}, see
also extended inflation \cite{La:1989za}, hybrid inflation
\cite{Linde:1990gz}\cite{Adams:1990ds} (ended by a first-order PT,
see
\cite{Cortes:2009ej}\cite{Lopez:2013mqa}\cite{Ashoorioon:2015hya}
for comparing it with the observations), chain inflation
\cite{Freese:2004vs}\cite{Freese:2005kt}\cite{Ashoorioon:2008pj}.

\begin{figure}[htbp]
\includegraphics[scale=2,width=0.47\textwidth]{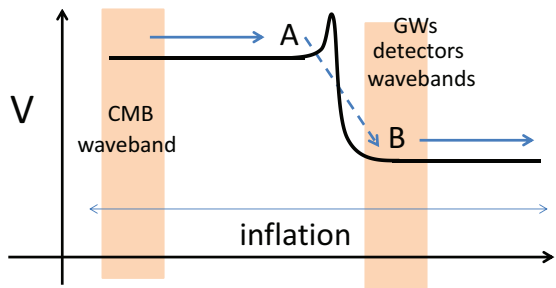}
\caption{Sketch of the inflation with first-order PT. The
realistic potential $ V(\phi,\psi)$ is similar to
that of hybrid inflation \cite{Linde:1990gz}\cite{Adams:1990ds}, where $\phi$ is the inflaton and $\psi$ is the scalar field responsible for the PT (see Appendix \ref{PT-app} for a model).
Initially, $\phi$ rolls slowly along its own direction (blue
solid line). At $t_*$, the $\psi$-direction PT (blue dash
line) completed quickly. Different from
\cite{Linde:1990gz}\cite{Adams:1990ds}, after $\rho_{released}$ is
diluted, the inflation will continue and $\phi$ will roll
slowly with a lower potential
$V(\phi,\psi_{B})<V(\phi,\psi_{A})$. }
    \label{fig-PT}
\end{figure}

Though currently the slow-roll inflation
\cite{Linde:1981mu}\cite{Albrecht:1982wi} is popular, the PT as
the birthright of inflation could always occur at certain time,
see Fig.\ref{fig-PT} (however, if the PT happened at the moment
corresponding to the CMB scale, it might be conflict with the
observations of CMB and large scale structure). During the
first-order PT, the lower-vacuum bubbles will nucleate with the
rate $\Gamma\sim e^{\beta(t-t_*)}$,  where $\beta^{-1}$ is approximately the duration of the PT. When ${\Gamma/ H^4}>\frac{9}{4\pi}$
\cite{Turner:1992tz}, the PT completed. The collisions of bubbles
will yield a stochastic GWs background with a peculiar spectrum
\cite{Kosowsky:1991ua}\cite{Caprini:2007xq}\cite{Huber:2008hg}.


We will calculate the primordial GWs spectrum brought by the
first-order PT during inflation, see Fig.\ref{fig-PT} for the
corresponding scenario. Tensor (GWs) mode $\gamma_{ij}$ satisfies
$\gamma_{ii}=0$, and $\partial_i \gamma_{ij}=0$, and the action of
$\gamma_{ij}$ is \be S^{(2)}=\int d\tau d^3x {M_p^2a^2\over
8}\lf[\lf(\frac{d\gamma_{ij}
}{d\tau}\rt)^2-(\vec{\nabla}\gamma_{ij})^2\rt]\,, \ee where
$\tau=\int dt/a$. We have set the propagating speed of GWs equal
to the speed of light, or see
\cite{Cai:2015yza}\cite{Cai:2016ldn}. The Fourier series of
$\gamma_{ij}$ is $\gamma_{ij}(\tau,\mathbf{x})=\int
\frac{d^3k}{(2\pi)^{3} }e^{-i\mathbf{k}\cdot \mathbf{x}}
\sum_{\lambda=+,\times} \hat{\gamma}_{\lambda}(\tau,\mathbf{k})
\epsilon^{(\lambda)}_{ij}(\mathbf{k})$,  where $
\hat{\gamma}_{\lambda}(\tau,\mathbf{k})=
\gamma_k(\tau)\hat{a}_{\lambda}(\mathbf{k}) +c.c.$,
$\epsilon_{ij}^{(\lambda)}(\mathbf{k})$ is the polarization
tensors, the annihilation and creation operators
$\hat{a}_{\lambda}(\mathbf{k})$ and
$\hat{a}^{\dag}_{\lambda}(\mathbf{k}^{\prime})$ satisfy $[
\hat{a}_{\lambda}(\mathbf{k}),\hat{a}_{\lambda^{\prime}}^{\dag}(\mathbf{k}^{\prime})
]=\delta_{\lambda\lambda^{\prime}}\delta^{(3)}(\mathbf{k}-\mathbf{k}^{\prime})$.
The energy density of GWs is \cite{Boyle:2005se} \be
\rho_{GW}=\sum_{\lambda=+,\times}\rho_{GW}^{\lambda}={ M_p^2\over
4}\int {k^3\over
2\pi^2}\lf({|{\gamma}^\prime_k|^2+{k^2}|\gamma_k|^2\over
a^2}\rt)d\ln{k}.\label{rho-GW}\ee While the equation of motion of
GWs mode $\gamma_k$ is \be
\frac{d^2u_k}{d\tau^2}+\left(k^2-\frac{a^{\prime\prime}}{a}
\right)u_k=0, \label{eom1} \ee where ${u}_k= {aM_p\over
2}\gamma_{k}$. Initially, the GWs modes should be deep inside
their horizon, i.e., $k^2 \gg \frac{a^{\prime\prime}}{a}$, so the
initial state is $u\sim \frac{1}{\sqrt{2k} }e^{-i k\tau}$.

When the PT happens, the collision of sub-horizon bubbles will
bring a sub-horizon GWs background, which will inevitably modify
the initial state of GWs modes. The energy spectrum of PT GWs is
given by \cite{Huber:2008hg}
\be
\Omega_{GW}^*(k)=\Omega_{GW}^{peak}{({\cal A}+{\cal B})k^{\cal B}_{peak}k^{\cal A}\over
{\cal B}k^{({\cal A}+{\cal B})}_{peak}+{\cal A}k^{({\cal A}+{\cal B})}}\,, \label{Omega-t}
\ee
\be
\Omega_{GW}^{peak}=\kappa^2 \lf({\Delta V_{inf}\over
V_{inf*}}\rt)^2\lf({H_{inf*}\over \beta}\rt)^2 {0.11v_b^3\over
0.42+v_b^2},
\ee
where ${\cal A}$ and ${\cal B}$ are exponents depicting the scale dependence of the spectrum such that $\Omega_{GW}^*(k)\sim k^{\cal A}$ for $k\ll k_{peak}$ and $\Omega_{GW}^*(k)\sim k^{-{\cal B}}$ for $k\gg k_{peak}$, $k_{peak}$ is the peak momentum of the spectrum; $\Delta V_{inf}=V_{inf*}-V_{inf}$,  $\kappa$ is the efficiency of converting vacuum energy into the bubble wall kinetic energy and
$\kappa\simeq 1$ for $V_{inf}\ll V_{inf*}$ (the subscript `$inf*$'
and `$inf$' correspond to the quantities at and after the PT,
respectively); the bubble is sub-horizon requires
$H_{inf*}/\beta<1$, and $v_b$ is the bubble wall velocity .

The sub-horizon GWs mode modified by the PT could be written as
\be u_k= {C(k)}e^{-i k\tau} \,,\label{uk1}\ee which is the
solution of Eq.(\ref{eom1}) for $k^2 \gg a^{\prime\prime}/a\simeq
a^2H^2$, where the information of PT is encoded in $C(k)$. When
$k^2 \gg a^2H^2$, Eq.(\ref{eom1}) suggests
$|\gamma_k^\prime|^2=(k^2+a^2H^2)|\gamma_k|^2\simeq
k^2|\gamma_k|^2$. Thus with Eqs.(\ref{rho-GW}) and (\ref{uk1}), we
have \be \Omega_{GW}^*={d\rho_{GW}\over \rho_{inf*}
\lf(d\ln{k}\rt)}={1\over 3\pi^2 M_p^2 H_{inf*}^2}{k^5\over
a^4_*}|C(k)|^2, \label{Omega1}\ee where
$\rho_{inf*}=3M_p^2H_{inf*}^2$. Thus we could get \be
|C(k)|^2={3\pi^2 M_p^2 H_{inf*}^2}{a^4_*\over k^5}\Omega_{GW}^*,\label{eq:Ck}
\ee where $\Omega_{GW}^*$ is set by (\ref{Omega-t}).  Hence, Eq.(\ref{eq:Ck}) is valid only
in the sub-horizon limit.

The inflation after the PT will occur with $\rho_{inf}\simeq
V_{inf}=V_{inf*}-\Delta V_{inf}$. We assumed that the energy
density $\rho_{released}\simeq \Delta V_{inf}$ deposited in bubble
walls has been efficiently released during the collisions of
bubbles
\cite{Watkins:1991zt}\cite{Kolb:1996jr}\cite{Zhang:2010qg}, and
diluted with the expansion of universe. The sub-horizon GWs (with
the wavelength $\lambda<H_{inf*}^{-1}\ll H_{inf}^{-1}$) brought by
the PT will be stretched outside the horizon $1/H_{inf}$.
By requiring that the solution of Eq.(\ref{eom1}) reduces to (\ref{uk1}) in the sub-horizon limit, we obtain $u_k({\tau})={-}C(k)\sqrt{-\pi k{\tau}\over 2}H^{(1)}_{3/2}(-k{\tau})$, which describes the perturbation
        modes both in sub- and super-horizon. On super-horizon
    scale, $ H^{(1)}_{3/2}(-k{\tau})\overset{-k{\tau}\rightarrow0}
    \approx-i \sqrt{2/(-\pi k^3\tau^3)}$, the spectrum $P_T$ of
    primordial PT GWs is \be P_T=\frac{4k^3}{\pi^2 M_p^2 a^2} \lf|u_k
    \rt|^2= {12 a_*^4 H^2_{inf*} H^2_{inf} \over k^4}\Omega_{GW}^*.
    \label{pt} \ee
The calculation is equivalent to that of
\cite{Jiang:2015qor}, so Eq.(\ref{pt}) has the same $k$-dependence
as that in \cite{Jiang:2015qor}. Thus we see that compared with
$P_{T,sub}$, $P_T\sim k^{-2}P_{T,sub}$ is reddened by inflation
($P_{T,sub}\sim k^3u_k^2\sim \Omega_{GW}^*/k^2$ is that for the
sub-horizon PT GWs, see also \cite{Jiang:2015qor}), which
thus records the unique imprint of inflation. The
information of the PT is encoded in $\Omega_{GW}^*$. It should be
mentioned that if the inflation ended after the first-order PT
\cite{Linde:1990gz}\cite{Adams:1990ds}, the PT GWs modes will stay
inside the horizon all the time, so $P_T=P_{T,sub}$ will preserve
the initial sub-horizon spectrum  and decrease with the
expansion of the universe as $P_{T,sub}\sim a^{-2}$, see
\cite{Lopez:2013mqa}.  We call the superhorizon GWs brought by the
PT (but stretched by inflation) the primordial PT GWs.

We have $v_b=1$ for $V_{inf}=V_{inf*}-\Delta V_{inf}\ll V_{inf*}$,
so that ${\cal A}\simeq 2.8$ and ${\cal B}\simeq 1$ in (\ref{Omega-t}). In
addition, the peak momentum of $\Omega_{GW}^*$ is $k_{peak}/(2\pi
\beta a_*)= 0.62/(1.8-0.1v_b+v_b^2)$ \cite{Huber:2008hg}, which
suggests $k_{peak}\simeq 1.4\beta {a_*}$ for $v_b\simeq 1$. Thus
Eq.(\ref{pt}) becomes \be P_T\simeq 0.24 \lf({V_{inf}\over
V_{inf*}}\rt)\lf(  \Delta V_{inf}\over V_{inf*}
\rt)^2\lf({H_{inf*}\over \beta}\rt)^6\cdot {3.8\lf({k\over
k_{peak}}\rt)^{-1.2}\over 1+2.8\lf({k\over k_{peak}}\rt)^{3.8}},
\label{pt1}\ee where $V_{inf*/inf}\simeq 3M_p^2H^2_{inf*/inf}$ is
used. We see that $P_T\sim k^{-5}$ for $k\gg k_{peak}$ is strongly
red, while $\sim k^{-1.2}$ for $k\ll k_{peak}$. The wavelength of
GWs mode brought by the PT is initially sub-horizon suggests a
cutoff frequency $k_{cutoff}={a_*}{H_{inf*}}<k$ for $P_T$. We have
$k_{peak}/k_{cutoff}\simeq 1.4\beta/H_{inf*}$, which is consistent
with the requirement of sub-horizon bubbles ($H_{inf*}/\beta<1$).

It is interesting to note that the maximal amplitude of
$\Omega_{GW}^*(k)$ brought by the PT is at $k_{peak}$, see
(\ref{Omega-t}), so it is with $P_{T,sub}$. However, the maximum
of $P_T$, i.e.,  $P_{T,max}$, is at $k= k_{cutoff}$, due to the
reddening effect of inflation. We have
\be P_{T,max}\simeq {1.4}
\lf( \Delta V_{inf}\over V_{inf*} \rt)^2 \lf({V_{inf}\over
V_{inf*}}\rt)\lf({H_{inf*}\over \beta}\rt)^{4.8} \label{pt2}
\ee
at
$k\simeq k_{cutoff}$ for  $(H_{inf*}/\beta)^{3.8}\ll1$. We
set ${V_{inf}/ V_{inf*}=0.1}$, which is consistent with the
requirement $V_{inf}\ll V_{inf*}$. Thus, if
${H_{inf*}/\beta}=0.1$, we have $P_{T,max}\sim {1.8}\times
10^{-6}$, while if ${H_{inf*}/\beta}=0.3$, $P_{T,max}\sim
{3.5}\times 10^{-4}$; in both cases,
$(H_{inf*}/\beta)^{3.8}\ll1$ is safely satisfied.  The
amplitude of $P_{T,max}$ depends on the choice of
$H_{inf}/\beta$. We see that $P_{T,max}$ could be far larger than
the power spectrum of the standard slow-roll inflation, which has
$P_T\lesssim 10^{-10}$.  However, it is also possible that
${H_{inf*}/\beta}< 0.013$, which indicates that
$P_{T,max}\lesssim 10^{-10}$ might even be smaller than the amplitude of the
GWs spectrum in the slow-roll regime.
Therefore, to generate observable PT GWs, it might require fine-tunning to some extent to obtain sufficiently large $H_{inf*}/\beta$ (see Appendix \ref{PT-app} for details).

\section{Detectability}


The signal-to-noise ratio (${\rm SNR}$) for the stochastic GWs
background is \cite{Thrane:2013oya} \be SNR=\sqrt{T_{\rm
obs}\int_{f_{\rm min}}^{f_{\rm max}}df
\frac{\Omega^2_{GW,PT}}{\Omega^2_{\rm sens}}}\,, \ee where
$T_{\text{obs}}$ is the total observation time, $[f_{\rm min},f_{\rm max}]$ defines the bandwidth of the detector,
$\Omega^{\rm LIGO}_{\rm sens}=\frac{S(f)}{\sqrt{2}\Gamma(f)}
\frac{2\pi^2}{3H_0^2}f^3$ and $\Omega^{\rm LISA}_{\rm sens}=S(f)
\frac{2\pi^2}{3H_0^2}f^3$ for LIGO~\cite{Allen:1997ad} and
LISA~\cite{Cornish:2001qi}, respectively. We assume the same noise
spectrum $S$ for the Livingston and Hanford detectors of LIGO,
with the overlap reduction $\Gamma(f)$~\cite{Thrane:2013oya}.
$\Omega_{GW,PT}(\tau_{0})=({1/\rho_c}){d\rho_{GW}(\tau_0)\over
d\ln{k}}$ is the energy spectrum of current GWs background, where
$\rho_{{c}}=3H^{2}_0/\big(8\pi G\big)$ is the critical energy density,
and $\rho_{GW}(\tau_0)$ is the energy density of GWs at present.
The spectrum $\Omega_{GW,PT}$ for primordial GWs background
(\ref{pt1}) is \cite{Turner:1993vb}
\begin{equation}
\label{density} \Omega_{GW,PT}(\tau_{0})=\frac{k^{2}}{12
a_0^2H^2_0}P_{T}(k)\lf[\frac{3
\Omega_{{m}}j_1(k\tau_0)}{k\tau_{0}}\sqrt{1.0+1.36\frac{k}{k_{\text{eq}}}
+2.50\big(\frac{k}{k_{\text{eq}}})^{2}}\rt]^2\,
\end{equation}
see also
e.g., \cite{Boyle:2005se}\cite{Zhao:2006mm}\cite{Kuroyanagi:2014nba}
for the details, where  $\Omega_m=\rho_m/\rho_c$, $j_1$ is the spherical-Bessel functions of order one, $\tau_0$ is the conformal time today,
$1/k_{eq}$ is the Hubble scale at matter-radiation equality.


\begin{figure}[htbp]
\includegraphics[scale=2,width=0.47\textwidth]{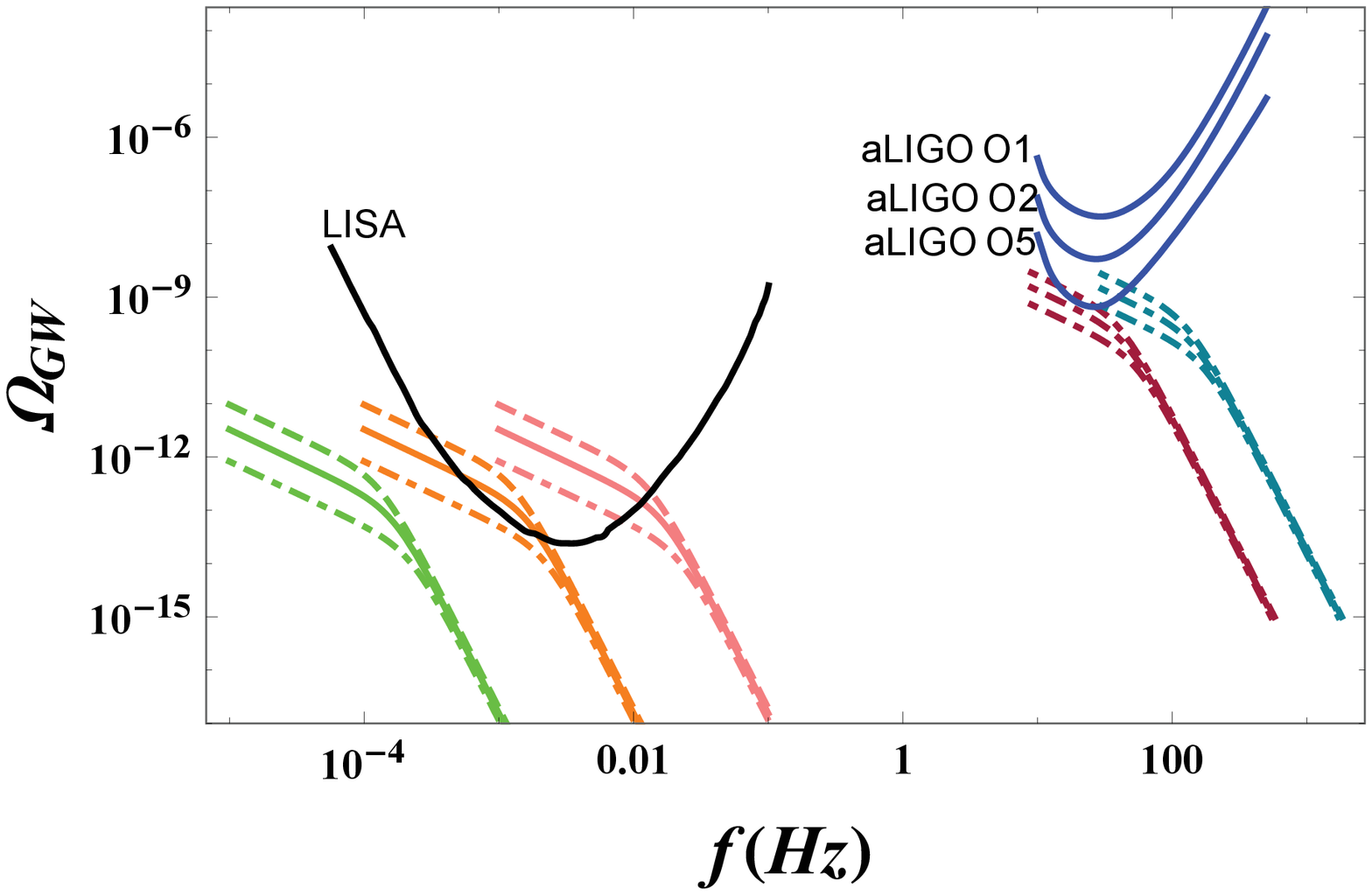}
\includegraphics[scale=2,width=0.47\textwidth]{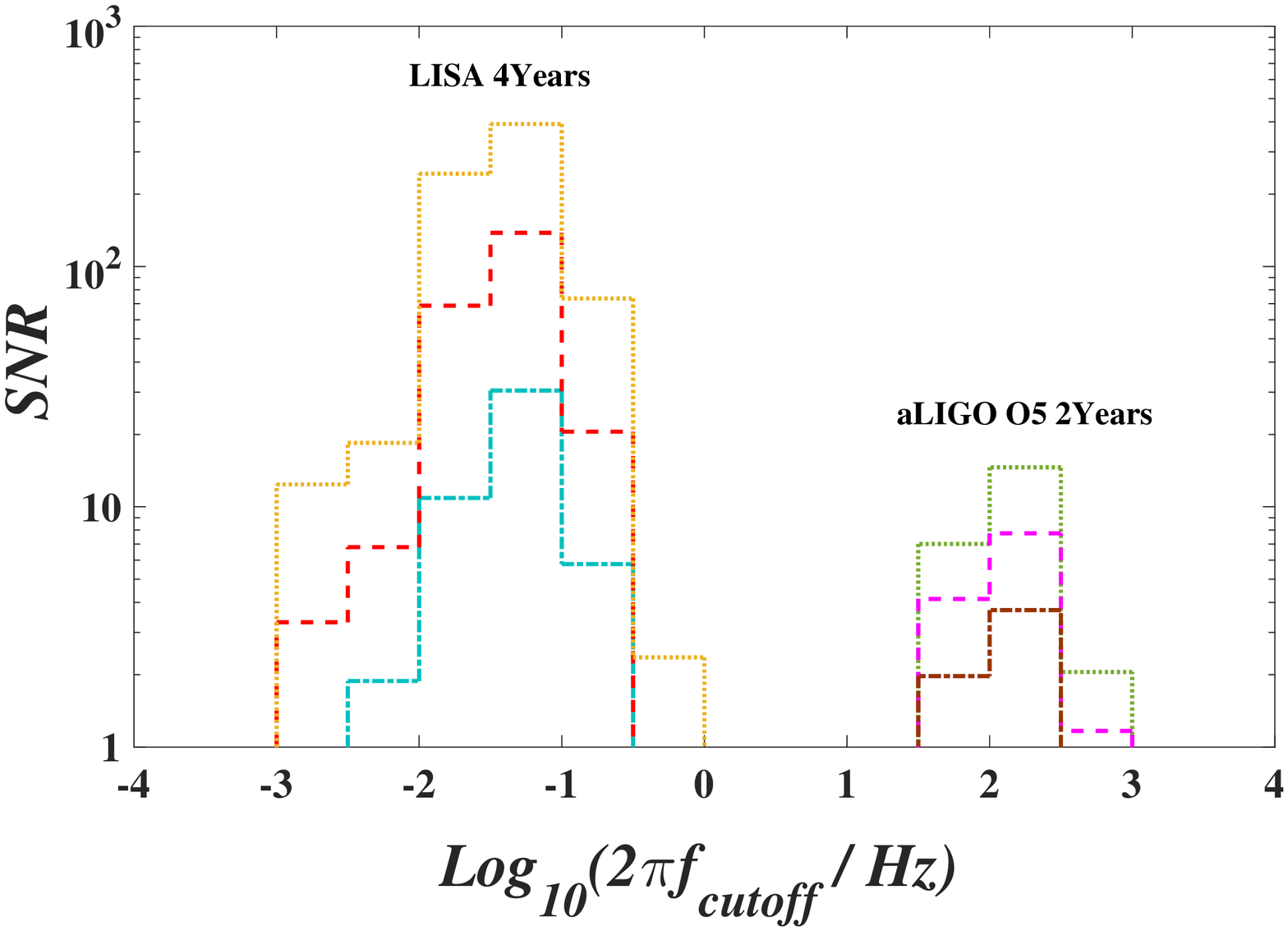}
\caption{Left panel: The stochastic background $\Omega_{GW,PT}$ in the
frequency bands of Advanced LIGO and LISA. In the LISA band, we set
the parameter ${H_{inf*}}/{\beta}= 0.075,\, 0.1,\, 0.125$,
respectively, for $f_{peak}\simeq10^{-4},\,10^{-3},\,10^{-2}$ Hz. In LIGO band, we set the parameter
${H_{inf*}}/{\beta}= 0.3,\, 0.35,\, 0.4$ for different $f_{peak}\simeq
100$Hz. Right panel: The SNR for $\Omega_{GW,PT}$ plotted in the
left panel. } \label{SNR}
\end{figure}

\begin{figure}[htbp]
\includegraphics[scale=2,width=0.47\textwidth]{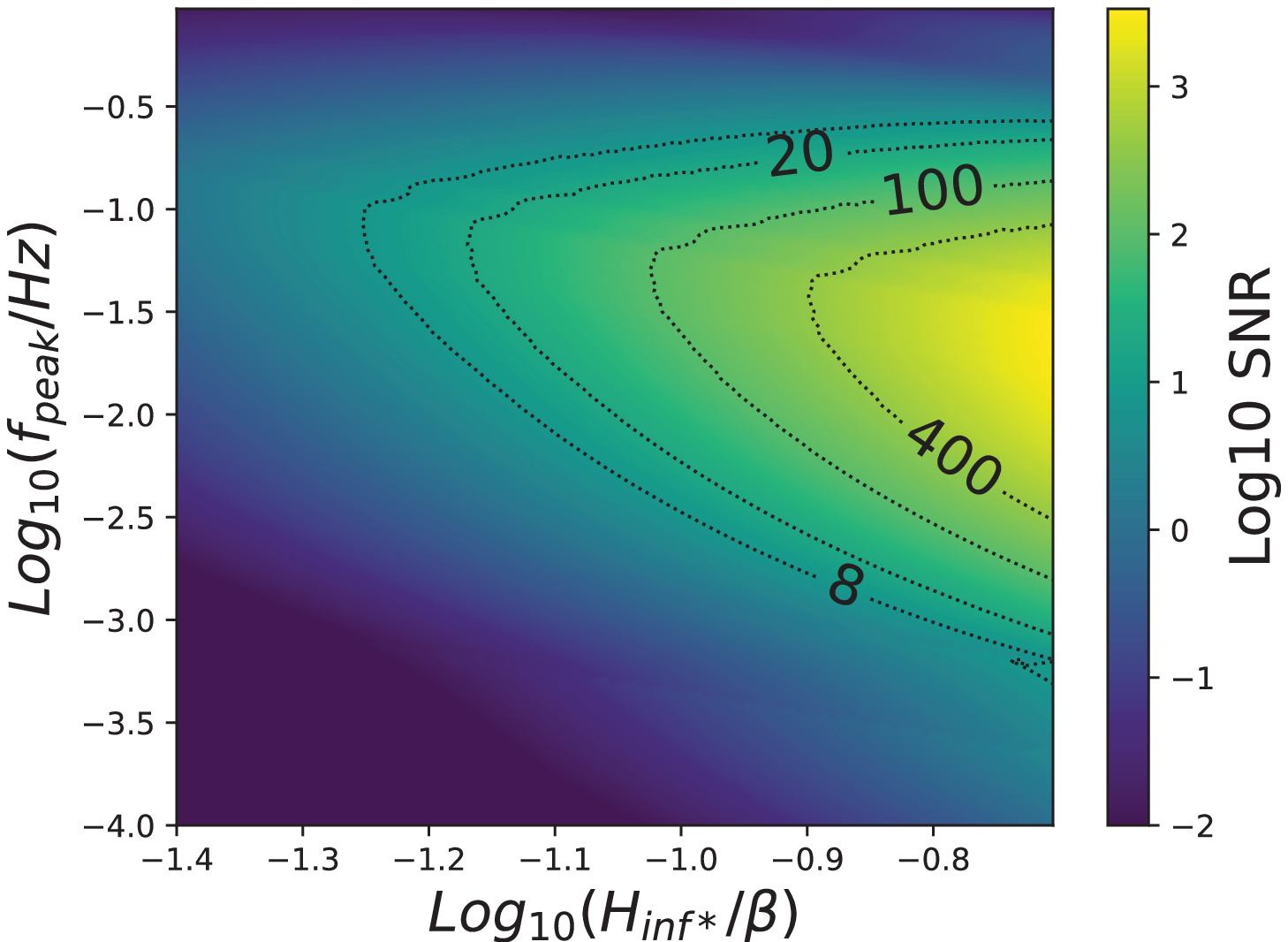}
\includegraphics[scale=2,width=0.47\textwidth]{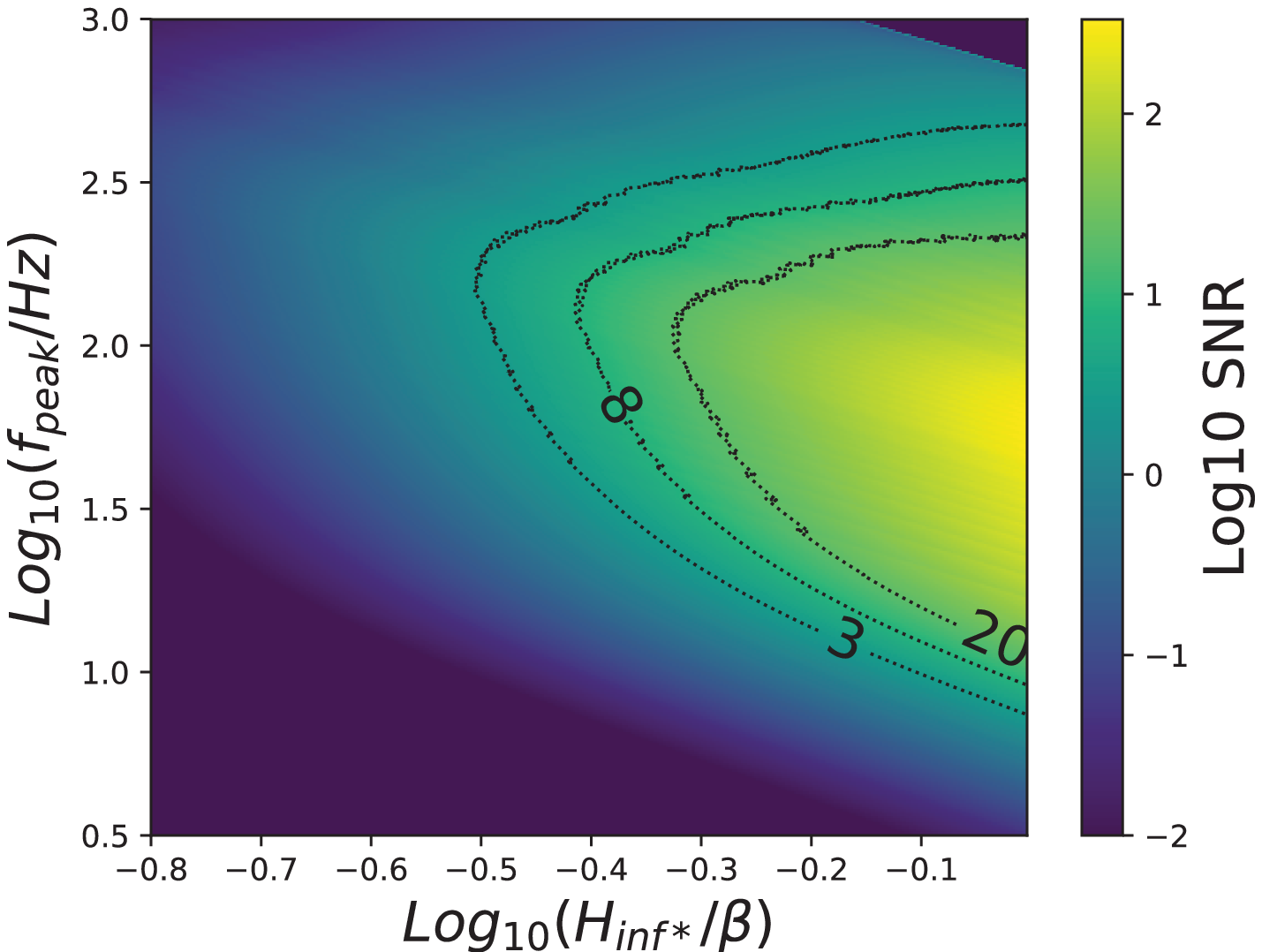}
\caption{Left panel: The contour of the SNR computed for LISA
with respect to the parameters $f_{peak}$ and
${H_{inf*}}/{\beta}$. Right panel: The contour of the SNR computed
for LIGO(O5).} \label{SNR2}
\end{figure}

The spectrum $\Omega_{GW,PT}(\tau_{0})$ is plotted in the left
panel of Fig.\ref{SNR}, and the corresponding SNR in the right
panel. $k_{peak}(f_{peak})$ and $H_{inf*}/\beta$ determine the
frequency band and magnitude of $\Omega_{GW,PT}$, respectively.
The stochastic background $\Omega_{GW,PT}$ with
$H_{inf*}/\beta\simeq 0.3$ at $f_{peak}\simeq 100Hz$ could be
marginally reached by LIGO(O5), while the signal of $\Omega_{GW,PT}$ with $H_{inf*}/\beta\lesssim
0.1$ at a lower frequency band $f_{peak}\sim 0.01Hz$ could be detected by LISA.
We plot the contours of the SNR computed for LIGO(O5) with 2-yrs
integration and the LISA with 4-yrs integration, respectively, in
Fig.{\ref{SNR2}}, which reflect the effects of the parameters
$f_{peak}$ and ${H_{inf*}}/{\beta}$ on the SNR.

We see that for $H_{inf}/\beta\gtrsim 0.3$, Advanced LIGO would
have the ability to find the signal of $\Omega_{GW,PT}$ with
SNR$\gtrsim 3$. However, the superpositions of GWs emitted by all
binary black holes (BBH) systems and other compact binaries will
also yield a stochastic GWs background $\Omega_{GW}\sim f^{2/3}$,
e.g., \cite{TheLIGOScientific:2016wyq}\cite{Abbott:2017xzg}. In
certain sense, this BBH background actually acts as a foreground.
To what extent can the primordial PT GWs background be
distinguished from the BBH foreground (once a stochastic signal is
 detected)?
To clarify this point, we will estimate the abilities of Advanced
LIGO to make a distinction between $\Omega_{GW,PT}$ and the BBH
stochastic GWs background $\Omega_{GW,PL}$, which has the
power-law behavior $\Omega_{GW,PL}=\Omega_0(f/f_0)^{2/3}$ (with
$f_0$ being a reference frequency) in the sensitive band of LIGO
\cite{TheLIGOScientific:2016wyq}.

We calculate the maximal likelihood ratio.  Following
\cite{Callister:2016ewt}, the likelihood ratio
\begin{equation}\label{likeratio} {\cal R}=\frac{{\cal
L}^{ML}(\Omega_{GW,PT}\vert\Omega_{GW,PT})}{{\cal
L}^{ML}(\Omega_{GW,PT}\vert\Omega_{GW,PL})}\ ,
\end{equation} where our $\Omega_{GW,PT}$ is regarded as the fiducial model, so
that ${\cal L}^{ML}(\Omega_{GW,PT}\vert\Omega_{GW,PT})={\cal N}$,
and the corresponding maximal likelihood for the power-law
background $\Omega_{GW,PL}$ is
\begin{equation}
\label{maxLikelihoodPL} {\cal
L}^{ML}(\Omega_{GW,PT}\,|\,\Omega_{GW,PL}) = {\cal N} \exp \left\{
-\frac{1}{4} \left( \left(\Omega_{GW,PT}\,|\,\Omega_{GW,PT}\right)
- \frac{
\left(\omega\,|\,\Omega_{GW,PT}\right)^2}{\left(\omega\,|\,\omega\right)}\right)\right\}
\end{equation} with the definitions $ (A\vert
B)=2T\left(\frac{3H_0^2}{2\pi^2}\right)^2\int_{f_{min}}^{f_{max}}
df\Gamma^2(f)\frac{A(f)B(f)}{f^6 S_1(f)S_2(f)}$
\cite{Allen:1997ad,Mandic:2012pj} and
$\omega(f)=\Omega_{GW,PL}/\Omega_0=(f/f_0)^{2/3}$. When the
likelihood ratio ${\cal R}\simeq 1$, both GWs backgrounds
($\Omega_{GW,PT}$ and $\Omega_{GW,PL}$) could be hardly
distinguished, while a fiducial GWs background could be identified
only for a large value of ${\cal R}$. Thus for the full
identifiability of the primordial PT GWs background, we must
require $\ln{\cal R}> 1$.

\begin{figure}[htbp]
\includegraphics[scale=2,width=0.48\textwidth]{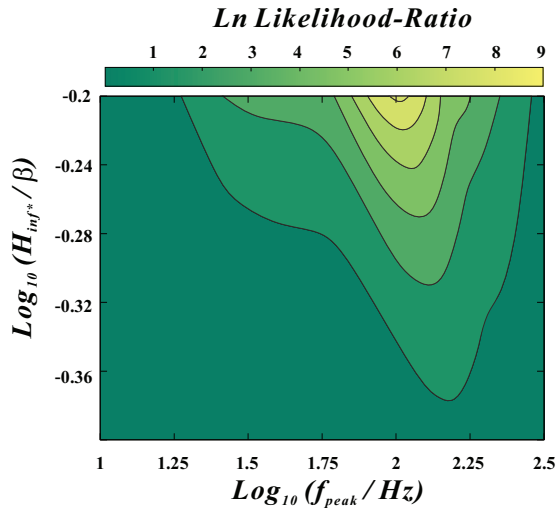}
\caption{Contour of the maximal log-likelihood ratio $\ln{\cal
R}$ given in (\ref{likeratio}) with respect to the parameters
${H_{inf*}}/{\beta}$ and $f_{peak}$.} \label{likelihood}
\end{figure}

The maximal log-likelihood ratio $\ln{\cal R}$ with respect to the
parameters ${H_{inf*}}/{\beta}$ and $f_{peak}$ is plotted in
Fig.\ref{likelihood}. We see that for LIGO O5, the region with
$H_{inf*}/\beta\gtrsim 0.3$ and $f_{peak}\simeq 100Hz$ could lead
to ${\cal R}\gg 1$, which corresponds to SNR$\gtrsim 3$ (see
Fig.\ref{SNR2}),
Note that the maximum of $\Omega_{GW,PT}$ is at $f_{cutoff}\simeq
f_{peak}/(1.4\beta/H_{inf*})$, as mentioned. Thus
$f_{cutoff}\simeq f_{peak}/4\simeq 25 Hz$ (for
$H_{inf*}/\beta\simeq 0.3$), which is just the most sensitive band
of LIGO. Thus if the parameter space $(f_{peak}, H_{inf*}/\beta)$
of primordial PT GWs is that plotted in Fig.\ref{SNR2}, the
Advanced LIGO at its design sensitivity would identify the
corresponding signal.


However, in the case of marginal sensitivity, it
would be difficult to prove the origin of the signal with respect
to other possible sources. Thus in order to link the signal
detected by GWs observatories to the inflation physics, a
multi-frequency detection is a key. More sensitive detectors
(the Einstein Telescope [45], the Cosmic Explore [46]) are
required. If $H_{inf*}/\beta<0.1$, we will have
$\Omega_{GW,PT}\ll \Omega_{GW,PL}$ in the frequency band of LIGO,
the BBH foreground $\Omega_{GW,PL}$ must be subtracted accurately
\cite{Regimbau:2016ike}, which is a challenging issue, since the
large uncertainty in the stellar BBH merger rate will inevitably
affect the amplitude of BBH GWs foreground
\cite{TheLIGOScientific:2016wyq}. Additionally, since the
reddened inflationary PT GWs spectrum satisfies
$\Omega_{GW}(k<k_{peak})\sim k^{-1.2}$ and
$\Omega_{GW}(k>k_{peak})\sim k^{-5}$, the $k<k_{peak}$ region
could be approximately mimicked by the $k>k_{peak}$ region
of the GWs spectrum due to a PT happened after inflation
($\Omega_{GW}(k>k_{peak})\sim k^{-1}$). Thus the signal should be
identified carefully with the multi-frequency detection before
linked to the inflation physics.

\section{Conclusion}

It is well-known that the first-order PT will yield a stochastic
GWs background with a peculiar spectrum. However, we
found that when such a PT happened during inflation, the GWs
spectrum brought by the PT will be reddened, which thus records
the unique voiceprint (the reddening-imprint) of inflation. We
estimated the abilities of Advanced LIGO and LISA to detect the
corresponding signal. Our result suggests that the GWs detectors
with higher sensitivity might consolidate our confidence that the
inflation has ever occurred.

The first-order PT is ubiquitous in the early universe. Though our
calculation is not sensitive to the details of model, the
corresponding model, e.g., along the line in
Refs.\cite{Linde:1990gz}\cite{Adams:1990ds}, is actually
interesting for study. Rich phenomenology caused by the PT
(during inflation), such as the primordial black holes
\cite{Kodama:1981gu}\cite{Garriga:2015fdk}\cite{Rubin:2000dq}, galactic nuclei \cite{Rubin:2001yw}\cite{Khlopov:2002yi}\cite{Khlopov:2004sc}, magnetic field, also
need to be fully understood, which might be also observable.
\\

\textbf{Acknowledgments}

YC would like to thank Burt Ovrut and Rehan Deen for hospitalities
and discussions during his visit at Penn. This work is
supported by NSFC, No.11575188, 11690021, and also supported by
the Strategic Priority Research Program of CAS, No.XDB23010100.
YTW is supported in part by the sixty-second batch of China
Postdoctoral Fund. YC is supported in part by the UCAS Joint PhD
Training Program.

\appendix

\section{A model}\label{PT-app}

In this Appendix, we will show a model with a first-order PT
happened during slow-roll inflation, slightly similar to that
proposed in the Appendix A of Ref. \cite{Liu:2009pk}.

The effective potential is a $(\phi,\psi)$-landscape \be
V= {\cal V}(\phi)+{\cal U}(\phi,\psi)\,,\ee  where \be
{\cal U}(\phi,\psi)={\sigma_1\over 16}\lf[\lf(\psi-{\mu\over
	\sqrt{\sigma_1}}\rt)^2-{\mu^2\over \sigma_1}\rt]^2-\sigma_2
{\sqrt{\sigma_1}\over 2\mu}\psi\,, \label{V1}\ee
$\sigma_1, \sigma_2=const.$, $\phi$ is the inflaton, $\psi$ is the scalar field responsible only for PT, $\mu$ depends on $\phi$ but is constant in the $\psi$-direction. Along the $\psi$-direction, ${\cal U}$ has the minima at around
$\psi\simeq 0$ (A) and $\psi\simeq {2\mu\over\sqrt{\sigma_1}}$ (B). The height and width of the barrier between both minima are about
$\mu^4/(16\sigma_1)$ and $2\mu/\sqrt{\sigma_1}$, respectively. Here, we
set \be \mu^2=m_\psi^2\lf[1+{\lambda_{\psi}\over
	M_p}\lf(\phi_{*}-\phi \rt)\rt], \label{mu2}\ee with
$m_{\psi}^2, \lambda_\psi =const.$ and $\phi_{*}$ being the PT point. As a result, the rolling of
$\phi$ along its potential ${\cal V}(\phi)$ will lower the
height of the barrier along $\psi$-direction.

\begin{figure}[htbp]
	\includegraphics[scale=2,width=0.5\textwidth]{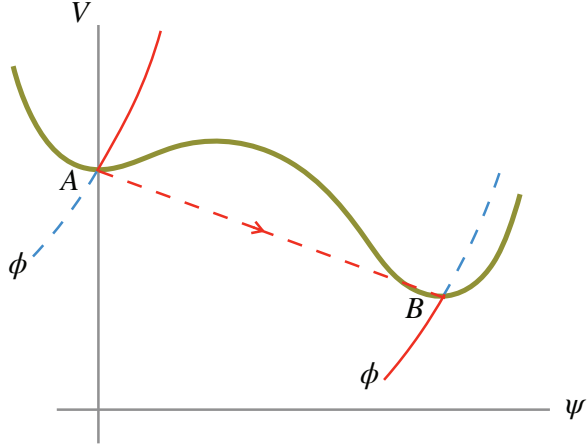}
	\caption{Sketch of our model. Initially, $\phi$ rolled along
		$V(\phi,\psi_{A})$ (red solid curve). Around $t_*$, the
		$\psi$-direction PT (red dashed curve) completed quickly. After the PT,
		$\phi$ will roll along a lower potential
		$V(\phi,\psi_{B})$ (right-lower red solid curve). }
	\label{fig-PT1}
\end{figure}

In the thin-wall approximation \cite{Coleman:1980aw}, the
nucleating rate of bubble is $\Gamma\sim e^{-B}$, where \be
B={27\pi^2 S_1^4\over 2\sigma_2^3}\simeq {\pi^2\mu^{12}\over
	6\sigma_1^4\sigma_2^3}.\ee According to Eq. (\ref{mu2}), defining $x=\pi^2
m_\psi^{12}/ (6\sigma_1^4\sigma_2^3)$, we have \be B\simeq
x\lf[1+{\lambda_{\psi}\over
	M_p}\lf(\phi_{*}-\phi \rt)\rt]^6\simeq
x\lf[1+{6\lambda_{\psi}\over
	M_p}\lf(\phi_{*}-\phi\rt)\rt]\label{B}\ee around
$\phi=\phi_*$. Thus $\Gamma\sim e^{-x}$ at
$\phi=\phi_*$. Dependent on the parameters $m_\psi, \sigma_1,
\sigma_2$, it is possible that when $\phi$ slowly rolled to the
PT point $\phi_*$, the PT along $\psi$-direction could complete
rapidly. After PT, the inflation will continue but $\phi$
will roll slowly along a lower potential $V(\phi,\psi_{B})<
V(\phi,\psi_{A})$, see Fig. \ref{fig-PT1} for a sketch.

We could estimate the bound for $H_{inf*}/\beta$, which appears in (\ref{pt2}). Considering
${\dot \phi}\simeq -{\cal V}^\prime/(3H)$ in the slow-roll
approximation in the $\phi$-direction, we have \be \phi_*-\phi\simeq
{-{\cal V}^{\prime} \over 3H}(t_*-t), \label{phistar} \ee
where we set ${\cal V}^{\prime}=\partial {\cal V}/\partial \phi=const.$ for convenience. Combining
Eqs. (\ref{B}) and (\ref{phistar}), we get
\be \Gamma\sim
	e^{-x}\exp\lf[{{2\lambda_\psi \lf|{\cal V}^{\prime}\rt|x\over
			M_p H }(t-t_*)}\rt]\sim e^{\beta(t-t_*)},
\ee
which suggests
$\beta={2\lambda_\psi \lf|{\cal V}^{\prime} \rt|x\over M_p H }$.
Thus, around the PT point, we have
\be {H_{inf*}\over \beta}\approx {M_p\over 6\sqrt{2} \lambda_{\psi}x\sqrt{\epsilon}}\,,
	\label{Hbeta}\ee
where $H_{inf*}$ is the Hubble parameter at
PT, and
$\epsilon=M_p^2({\cal V}^{\prime} /{\cal V})^2/2$ is the slow-roll
parameter along the $\phi$-direction, which should also be evaluated at the PT point. Generally, $\lambda_{\psi}<1$. According to (\ref{Hbeta}),
it is possible to have $H_{inf*}/\beta\simeq 0.1$ provided
$\lambda_\psi$, $x$ and $\epsilon$ are sufficiently small.

However, it should be mentioned that if the PT is not completed at
$\phi=\phi_*$, after PT, some regions will inflate along
$V(\phi,\psi_{A})$, while others along the lower potential
$V(\phi,\psi_{B})$, see multi-stream inflation
\cite{Li:2009sp}\cite{Li:2009me}. The relevant issue is also
interesting for study.

 \end{document}